\documentclass[conference]{IEEEtran}
\IEEEoverridecommandlockouts
\usepackage[T1]{fontenc}
\usepackage{amsmath}
\usepackage{amsfonts}
\usepackage{graphicx}
\usepackage{color}

\usepackage{amsmath}
\usepackage{cite}
\usepackage{amsfonts}
\usepackage{amssymb}
\usepackage{graphicx}
\usepackage{subfigure}
\usepackage{algorithm}
\usepackage{algpseudocode}
\usepackage{mathtools}

\usepackage{textcomp}
\usepackage[export]{adjustbox}
\usepackage{float}
\usepackage{booktabs}
\usepackage{multicol}
\usepackage{xparse}
\usepackage{color,soul}
\usepackage[bottom]{footmisc}
\usepackage[binary-units=true]{siunitx}
\DeclareSIUnit{\belmilliwatt}{Bm}
\DeclareSIUnit{\dBm}{\deci\belmilliwatt}
\DeclareSIUnit[per-mode=symbol,per-symbol=p]{\Bps}{\byte\per\second}
\sethlcolor{yellow}
\usepackage{algpseudocode}

\makeatletter
\def\BState{\State\hskip-\ALG@thistlm}
\makeatother

\usepackage{suffix,mathtools}
\DeclarePairedDelimiterX\MeijerM[3]{\lparen}{\rparen}%
{#3\,\delimsize\vert\begin{smallmatrix}#1 \\ #2\end{smallmatrix}}
\newcommand\MeijerG[8][]{%
  G^{\,#2,#3}_{#4,#5}\MeijerM[#1]{#6}{#7}{#8}}

\WithSuffix\newcommand\MeijerG*[7]{%
  G^{\,#1,#2}_{#3,#4}\MeijerM*{#5}{#6}{#7}}

\def\BibTeX{{\rm B\kern-.05em{\sc i\kern-.025em b}\kern-.08em
    T\kern-.1667em\lower.7ex\hbox{E}\kern-.125emX}}

\title{Physical Layer Security in Finite Blocklength Massive IoT with Randomly Located Eavesdroppers

\author{
\IEEEauthorblockN{Tijana Devaja\IEEEauthorrefmark{1}, Milica Petkovic\IEEEauthorrefmark{1}, Sokol Kosta\IEEEauthorrefmark{2}, Dejan Vukobratovi\' c\IEEEauthorrefmark{1}, and \v Cedomir Stefanovi\' c\IEEEauthorrefmark{2}}
\IEEEauthorblockA{
\IEEEauthorrefmark{1}Faculty of Technical Sciences, University of Novi Sad, Serbia (\{tijana.devaja, milica.petkovic, dejanv\}@uns.ac.rs)\\
\IEEEauthorrefmark{2}Department of Electronic Systems, Aalborg University, Aalborg, Denmark (\{sok,cs\}@es.aau.dk)}}

\thanks{This article/publication is based upon work from COST Action 6G-PHYSEC (CA22168), supported by COST  (European Cooperation in Science and Technology). This work has received funding from Serbian Ministry of Science, Technological Development and Innovation, through the Science and Technological Cooperation program Serbia-China, Research and development project No 00101957 2025 13440 003 000 620 021.}
}

\begin{document}
\maketitle

\begin{abstract}
This paper analyzes the physical layer security performance of massive uplink Internet of Things (IoT) networks operating under the finite blocklength (FBL) regime. IoT devices and base stations (BS) are modeled using a stochastic geometry approach, while an eavesdropper is placed at a random location around the transmitting device. This system model captures security risks common in dense IoT deployments. Analytical expressions for the secure success probability, secrecy outage probability and secrecy throughput are derived to characterize how stochastic interference, fading and eavesdropper spatial uncertainty interact with FBL constraints in short packet uplink transmissions. Numerical results illustrate key system behavior under different network and channel conditions.
\end{abstract}

\section{Introduction}

Massive Internet of Things (IoT) networks are one of the key components of  modern wireless communications, enabling large-scale connectivity among heterogeneous low-power devices \cite{5g,mag,lpwan}. In many practical scenarios, IoT nodes transmit short status updates or control packets, often under strict latency and reliability constraints. Such scenarios cannot be accurately captured by the classical Shannon capacity framework, since it relies on blocklengths that are practically infinite and therefore not suitable for short, latency-limited packets. Consequently, finite blocklength (FBL) information theory has emerged as a more accurate tool for evaluating the performance of short packet transmissions, providing tight approximations for decoding error probability in realistic IoT environments \cite{Polyanskiy2010,pkd2016,Durisi15,durisi}.

Massive IoT deployments are exposed to various security threats, with eavesdropping being one of the most fundamental. This issue becomes even more critical when IoT devices operate under  energy, latency  or hardware constraints, which often limits traditional cryptographic solutions. Physical layer security (PLS) provides a complementary approach by exploiting randomness in noise, fading and network geometry to ensure  secure communications at the signal level. Although PLS has been extensively studied in large-scale wireless networks using stochastic geometry, most existing works rely on the idealized infinite blocklength regime and therefore fail to capture the fundamental reliability and secrecy tradeoffs in short packet communications  \cite{rad1, rad2, rad3}.

Despite the growing interest in both FBL communications and PLS, secure performance in massive IoT systems remains underexplored when all network entities, including potential adversaries, follow spatially random locations typically modeled through stochastic geometry. Although there are recent studies on PLS performance in the FBL regime \cite{PLS_FBL_1,PLS_FBL_2,PLS_FBL_3,PLS_FBL_4}, to the best of the authors' knowledge, only a few existing works consider systems in which the eavesdropper is modeled as a spatially random node under a stochastic-geometry framework in the presence of FBL constraints \cite{PLS_FBL_PPP_1,PLS_FBL_PPP_2}. More importantly, to date there are no works analyzing PLS performance in the FBL regime for massive IoT networks in which both IoT nodes and base stations (BSs) follow spatial randomness.

Motivated by this research gap, this paper analyzes a massive IoT uplink network in which BSs and IoT devices are modeled as independent Poisson point processes (PPPs), while an eavesdropper is uniformly distributed within a disk of a certain radius around a transmitting IoT device. This model captures scenarios where an eavesdropper is located in the physical proximity of the targeted IoT device, representing local security threats common in massive IoT deployments. The analysis is conducted in the FBL regime, enabling accurate characterization of decoding error probabilities at both the intended BS and the eavesdropper. No prior work has examined the joint impact of stochastic interference, short-packet reliability, and the spatial uncertainty of the eavesdropper within a unified analytical framework. The analytical expressions for secure success probability, secrecy outage probability  and secrecy throughput  are derived and used to obtain numerical results in order to illustrate main system behavior under different system and channel parameters. 

The rest of the paper is organized as follows. Section~II introduces the system and channel model. Section~III presents the FBL framework, while Section~IV derives secrecy performance metrics under FBL.  Section~V provides numerical results and discussion, while Section~VI concludes the paper.

\section{System and Channel Model}
\label{sec:system_model}

\subsection{Network Geometry and Random Access Scheme}

\begin{figure}[!t]
\centerline{\includegraphics[width=3.5in,height=2in]{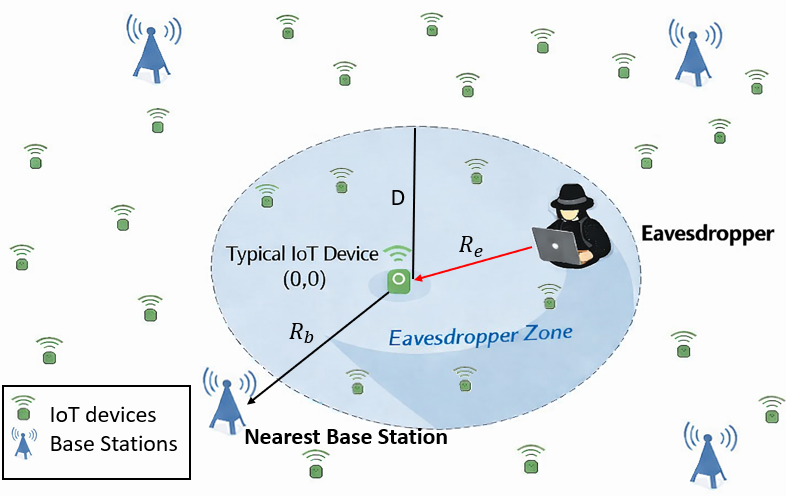}}
\caption{System model.}
\label{Fig1}
\end{figure}

We consider an uplink massive IoT cellular network that connects a set of IoT devices deployed over $\mathbb{R}^2$ plane, as depicted in Fig.~\ref{Fig1}. The IoT devices access the BSs using the slotted ALOHA protocol, with communication taking place in equal-length time slots. IoT devices are distributed according to a  homogeneous PPP $\Phi_u$ with density $\lambda_u$ (devices/m$^2$), while BSs are distributed according to an independent homogeneous PPP $\Phi_b$ with density $\lambda_b$ (BSs/m$^2$). This spatial modeling approach is standard in the analysis of large-scale wireless networks and massive IoT systems \cite{Haenggi12,Novlan13}.
In each slot each device independently accesses the channel with
probability $p$, resulting in an active PPP
$\Phi_a \subset \Phi_u$ with density $\lambda_a = p\lambda_u$.
This baseline random-access massive IoT model follows
\cite{tijana,Hesham,Hesham1}. To capture potential security threats, we also introduce a passive eavesdropper positioned at a random position around the typical IoT device, uniformly distributed within a disk of radius $D$, as presented in Fig.~\ref{Fig1}. 

\subsection{Association and Distance Distributions}

Each active IoT device associates with its geographically nearest BS. For a typical active device, let $b_0 \in \Phi_b$ denote its serving BS, and let $R_b$ be the corresponding distance. Under nearest-BS association in a PPP deployment (where the serving BS acts as Bob in the PLS framework), the probability density function (PDF) of $R_b$ is given by \cite{Haenggi12}
\begin{equation}
    f_{R_b}(r_b)=2\pi\lambda_b r_b \exp\!\big(-\pi\lambda_b r_b^2\big), \qquad r_b>0,
    \label{eq:fRb}
\end{equation}
which captures the well-known nearest-neighbor distribution in a homogeneous PPP.

\subsection{Eavesdropper Location Model}
To incorporate the PLS aspect into considered massive IoT framework,  we introduce a single passive eavesdropper (acting as Eve) whose goal is to intercept the uplink transmission of the typical active IoT device.
The eavesdropper is positioned at a random location around the transmitting device, uniformly distributed within a disk of radius $D$.
We denote by $R_e$ the distance between the typical IoT device and the eavesdropper, whose PDF is
\begin{equation}
f_{R_e}(r_e)=\frac{2r_e}{D^2},\quad 0\le r_e \le D.
\label{eq:fRe}
\end{equation}

\subsection{Signal-to-Interference Ratio (SIR)}

All IoT devices transmit with normalized unit power. Assuming an interference-limited regime, which is typical for uplink massive IoT networks, the instantaneous SIR at both the serving BS (Bob) and the eavesdropper (Eve) is defined as
\begin{equation}
    \gamma_{\star} = \frac{G_{\star} H_{\star} R_{\star}^{-\eta}}{I_{\star}},
    \label{eq:sinr_b}
\end{equation}
where the subscript $\star \in \{b,e\}$ is used as a generic index to distinguish between Bob and Eve.\footnote{Throughout the paper, the symbol $\star \in \{b,e\}$ denotes either the legitimate receiver (Bob) or the eavesdropper (Eve).}
The small-scale fading power gains of the channels toward the BS and the eavesdropper are denoted by $H_b$ and $H_e$, respectively, while the corresponding aggregated interference terms are denoted by $I_b$ and $I_e$. 
Large-scale path-loss is modeled as $R_{\star}^{-\eta}$, where $\eta>2$ is the path-loss
exponent. The distances $R_b$ and $R_e$ are previously defined with the PDFs given by \eqref{eq:fRb} and \eqref{eq:fRe}, respectively.
In addition, $G_b$ and $G_e$ represent the receiver antenna gains at the BS and at the eavesdropper, respectively, which are  not identical due to different hardware capabilities.
The BS is equipped with sectorized or directional antennas that provide a higher receiver
gain, while IoT-class devices employ compact omnidirectional antennas with limited gain.

\subsection{Small-Scale Fading Model}

Small-scale fading is modeled by the Nakagami-$m$ distribution, thus the fading power gain follows a Gamma distribution with the PDF given as \cite{simon}
\begin{equation}
f_{H_{\star}}(h_{\star})=\frac{m^{m}}{\Gamma(m)} h_{\star}^{\,m-1} e^{-m h_{\star}}, \qquad h_{\star}>0,
\label{eq:nakagami}
\end{equation}
where $\mathbb{E}[H_{\star}]=1$, and $m \ge 1$ denotes the Nakagami-$m$ fading parameter. The same fading severity (i.e., identical parameter $m$) is assumed for both the IoT–BS and IoT–eavesdropper links, while fading realizations remain independent across links. The Nakagami-$m$ model encompasses a wide range of fading conditions and is commonly adopted in massive IoT and cellular network analyses due to its analytical tractability.

\subsection{Interference Field and Statistics}

All IoT devices are spatially distributed according to a homogeneous PPP
$\Phi_u\subset\mathbb{R}^2$.
Due to random access, only a fraction of devices are active at a given time,
resulting in a thinned PPP of interferers.
The aggregate uplink interference observed at receiver $\star\in\{b,e\}$ is
\begin{equation}
I_{\star}=\sum_{i\in\Phi_u} a_i h_{i,\star} r_{i,\star}^{-\eta},
\label{eq:interference_star}
\end{equation}
where $a_i\in\{0,1\}$ denotes the activity indicator, $h_{i,\star}$ is the fading
coefficient, $r_{i,\star}$ is the distance between the $i$-th interferer and
receiver $\star$ (being $b$ for the legitimate receiver, i.e., the nearest BS, or $e$ for the eavesdropper).

The distribution of the aggregate interference is characterized via its Laplace transform
(LT).
For PPP devices with slotted ALOHA access and Nakagami-$m$ fading, the LT of the
interference power has the Kohlrausch--Williams--Watts (KWW) form \cite{R6}
\begin{equation}
\mathcal{L}_{I_{\star}}(s)
=
\mathbb{E}\!\left[e^{-s I_{\star}}\right]
=
\exp\!\left(-t\, s^{\frac{2}{\eta}}\right),
\label{eq:LI}
\end{equation}
where
\begin{equation}
t= p\lambda_u \pi\,
\frac{\Gamma\!\left(m+\tfrac{2}{\eta}\right)}{\Gamma(m)\,m^{\frac{2}{\eta}}}\,
\Gamma\!\left(1-\tfrac{2}{\eta}\right).
\label{eq:t_param}
\end{equation}
The same interference statistics apply at both the legitimate receiver and the
eavesdropper, since interference is generated by the same PPP of active IoT devices.

The interference PDF can be obtained via inverse Laplace
transform (ILT).
In the special case $\eta=4$, the interference follows a Lévy distribution \cite{R6}
\begin{equation}
f_{I_{\star}}(x)
=
\frac{t \exp\!\left(-\tfrac{t_{\star}^2}{4x}\right)}{2\sqrt{\pi}\,x^{3/2}}.
\label{eq:levy_I}
\end{equation}
To perform the FBL  analysis, we derive the conditional PDF of SIR $\gamma_{\star}$  defined in \eqref{eq:sinr_b}, conditioned on $h_{\star}$ and $r_{\star}$. For $\eta = 4$ and using the L\'evy form of $f_{I_{\star}}$ in \eqref{eq:levy_I}, the conditional PDF of SIR  is obtained as
\begin{equation}
f_{\Gamma_{\star}}(\gamma_{\star}|h_{\star},r_{\star})
= \frac{t\exp\!\big(-\tfrac{t^2 \gamma_{\star}}{4G_{\star}h_{\star} r_{\star}^{-4}}\big)}{2\sqrt{\pi G_{\star}h_{\star} r_{\star}^{-4}\gamma_{\star}}}.
\label{pdfSINR}
\end{equation}

\section{Finite Blocklength Decoding Model}
\label{sec:fbl_model}

We consider short packet transmission, where codewords of length $n$ channel uses are transmitted at a coding rate
$R$ bits per channel use.
The block error probability conditioned on an instantaneous SIR, denoted by $\gamma_{\star}$ ($\star \in \{b,e\}$), under FBL coding is well approximated by the
normal approximation \cite{Polyanskiy2010,pkd2016}
\begin{equation}
\epsilon_{\star}(\gamma_{\star})
\approx
Q\!\left(
\sqrt{\frac{n}{V(\gamma_{\star})}}
\big(C(\gamma_{\star})-R\big)
\right),
\label{eq:fbl_eps}
\end{equation}
where $C(\gamma_{\star})=\log_2(1+\gamma_{\star})$ is the Shannon capacity and $
V(\gamma_{\star})
= \gamma_{\star}(\gamma_{\star}+2)(\log_2 e)^2 /(1+\gamma_{\star})^2
$
denotes the channel dispersion.
Following the standard practice in stochastic-geometry-based FBL analysis, aggregate
interference is treated as Gaussian noise \cite{durisi,Hesham}.

\subsection{Block Error Probability at the Legitimate Receiver}

Using the SIR $\gamma_b$ at the serving BS (Bob), the unconditional uplink
block error probability is determined as \cite{tijana}
\begin{equation}
\begin{split}
& \epsilon_b
=  \int_{0}^{\infty}
\!\!\epsilon_b(\gamma_b)\,
f_{\Gamma_b}\!(\gamma_b) {\rm d}\gamma_b\\
&=\!\!\!
\int_{0}^{\infty}\!\!\!\!\int_{0}^{\infty}\!\!\!\!\int_{0}^{\infty}
\!\!\!\!\!\!\!\epsilon_b(\gamma_b)\,
f_{\Gamma_b}\!(\gamma_b \!\mid \!h_b,r_b)\,\!
f_{H_b}(h_b)\,\!
f_{R_b}(r_b)\,\!
{\rm d}\gamma_b\, \!{\rm d}h_b\, {\rm d}r_b,
\end{split}
\label{eq:UL_Pe_full}
\end{equation}
where $f_{R_b}(r_b)$,  $f_{H_b}(h_b)$ and $f_{\Gamma_b}(\gamma_b \!\mid \!h_b,r_b)$ are the  PDFs defined in \eqref{eq:fRb}, \eqref{eq:nakagami} and \eqref{pdfSINR}, respectively.

Since the integral in \eqref{eq:UL_Pe_full} can not be derived in a closed-form, we
employ the linear approximation of the $Q$-function with the parameters $\mu=\sqrt{\frac{n}{2\pi(2^{2R}-1)\log_2^2e}}$, $\theta=2^R-1$ and $a=\sqrt{\frac{\pi}{2\mu^2}}$.
Following the derivation in \cite{tijana}, the uplink block error probability at the
BS (Bob) is tightly approximated as
\begin{equation}
\begin{split}
& \epsilon_b
\approx\!
\frac{a_1}{\Gamma(m)\pi}
\MeijerG*{2}{3}{3}{3}{1, 0, \frac{1}{2}}{\frac{1}{2}, m, 0}{\!z_1\!}
- b_1 z_1 \!\left(c + d z_1^{m-\frac{1}{2}}\right)
 \\
& +
\frac{a_2}{\Gamma(m)\pi}
\MeijerG*{2}{3}{3}{3}{1, 0, \frac{1}{2}}{\frac{1}{2}, m, 0}{\!z_2\!}
+ b_2 z_2 \!\left(c + d z_2^{m-\frac{1}{2}}\right),
\end{split}
\label{eq:Peb_fbl}
\end{equation}
where  
$a_1 = \frac{1}{2} + \frac{\mu\theta}{\sqrt{2\pi}}$, 
$a_2 = \frac{1}{2} - \frac{\mu\theta}{\sqrt{2\pi}}$, $b_1 = \frac{\mu(\theta + a)}{\sqrt{2\pi}}$, 
$b_2 = \frac{\mu(\theta - a)}{\sqrt{2\pi}}$,  $z_1 = \frac{t^2(\theta + a)m}{G_b( \lambda_b \pi)^2}$,  $z_2 = \frac{t^2(\theta - a)m}{G_b(\lambda_b \pi)^2}$, $c = \frac{\Gamma(m - \frac{1}{2})}{3\sqrt{\pi} \Gamma(m)}$, $d = \frac{(-1)^{m} m}{m+1}$, and $\MeijerG*{\cdot}{\cdot}{\cdot}{\cdot}{\cdot}{\cdot}{\cdot}$ is the Meijer G-function \cite[(9.301)]{grad}.

\subsection{Block Error Probability at the Eavesdropper}

The eavesdropper  attempts to decode the same codeword at rate $R$ under FBL constraints.
Conditioned on the instantaneous SIR denoted by $\gamma_e$ defined in \eqref{eq:sinr_b}, the block error probability under FBL at Eve is
 given by $\epsilon_e(\gamma_e)$ in \eqref{eq:fbl_eps}.
Differently to the legitimate receiver, the eavesdropper location is uniformly
distributed within a disk of radius $D$ centered at the typical IoT device, resulting
in a different SIR distribution.

Accordingly, the unconditional block error probability at the Eve is
\begin{equation}
\begin{split}
& \epsilon_e
=  \int_{0}^{\infty}
\!\!\epsilon_e(\gamma_e)\,
f_{\Gamma_e}\!(\gamma_e) {\rm d}\gamma_e\\
&=\!\!\!
\int_{0}^{D}\!\!\!\!\int_{0}^{\infty}\!\!\!\!\int_{0}^{\infty}
\!\!\!\!\!\!\!\epsilon_e(\gamma_e)\,
f_{\Gamma_e}\!(\gamma_e \!\mid \!h_e,r_e)\,\!
f_{H_e}(h_e)\,\!
f_{R_e}(r_e)\,\!
{\rm d}\gamma_e\, \!{\rm d}h_e\, {\rm d}r_e,
\end{split}
\label{eq:UL_Pe_full1}
\end{equation}
where $f_{R_e}(r_e)$,  $f_{H_e}(h_e)$ and $f_{\Gamma_e}(\gamma_e \!\mid \!h_e,r_e)$ are the  PDFs defined in \eqref{eq:fRe}, \eqref{eq:nakagami} and \eqref{pdfSINR}, respectively.

Using the same linear approximation of the $Q$-function as for Bob \cite{tijana}, after following derivation summarized in Appendix A, the block error
probability at Eve is tightly approximated as
\begin{equation}
\begin{split}
\epsilon_e
&\approx
B\bigg (
\MeijerG*{2}{5}{5}{7}{-\frac{1}{4}, 0, \frac{1}{4},\frac{1}{2}, 1}{\frac{1}{2}, m, 0, -\frac{1}{2}, -\frac{1}{4}, 0, \frac{1}{4}}{{y}_{1}} \\
& + A \MeijerG*{2}{5}{5}{7}{-\frac{1}{4}, 0, \frac{1}{4},\frac{1}{2}, 1}{\frac{1}{2}, m, 0, -\frac{1}{2}, -\frac{1}{4}, 0, \frac{1}{4}}{{y}_{2}}\bigg)\\
 &+C_2 \MeijerG*{2}{5}{5}{7}{-\frac{3}{4}, -\frac{1}{2}, -\frac{1}{4},0,-\frac{1}{2}}{-\frac{3}{2},0, m-\frac{1}{2}, -1, -\frac{3}{4}, -\frac{1}{2}, - \frac{1}{4}}{{y}_{2}} \\
 & - C_1 \MeijerG*{2}{5}{5}{7}{-\frac{3}{4}, -\frac{1}{2}, -\frac{1}{4},0,-\frac{1}{2}}{-\frac{3}{2},0, m-\frac{1}{2}, -1, -\frac{3}{4}, -\frac{1}{2}, - \frac{1}{4}}{{y}_{1}},
\end{split}
\label{eq:Pee_fbl_closedform}
\end{equation}
where $y_{1}=\frac{D^4t^2(\theta + a)m}{4G_e},
y_{2}=\frac{D^4t^2(\theta - a)m}{4G_e},$  $A = \frac{1}{2} - \frac{\mu\theta}{\sqrt{2\pi}}$, $B=\frac{2D}{\eta \sqrt{\pi}\Gamma(m)}$, $C_{1}=\frac{D^3t\mu\sqrt{m}(\theta + a)^{\frac{3}{2}}}{\sqrt{2}\pi \Gamma(m)\eta}$, $C_{2}=\frac{D^3t\mu\sqrt{m}(\theta - a)^{\frac{3}{2}}}{\sqrt{2}\pi \Gamma(m)\eta}$. 

\section{Secrecy Metrics Under Finite Blocklength}
\label{sec:secrecy_metrics_fbl}

In short-packet massive IoT systems, secrecy is inherently linked to the decoding outcomes
at both the legitimate receiver and the eavesdropper.
Unlike classical information-theoretic secrecy metrics based on asymptotic capacity,
FBL communication exhibits a smooth transition between successful
and unsuccessful decoding.
Motivated by this observation, we adopt a decoding-based secrecy formulation,
which has recently emerged as a meaningful operational security metric for short-packet
wireless systems \cite{Mukherjee2014}.

\subsection{Finite Blocklength Secure Success Probability}
\label{subsec:fbl_secure_success}

A transmission is said to be securely successful if the legitimate BS
Bob decodes the packet correctly, while the eavesdropper Eve fails to decode it, i.e., 
the secure success event is defined as
\begin{equation}
\mathcal{S} \triangleq \{\text{Bob decodes}\} \cap \{\text{Eve fails}\}.
\end{equation}
Decoding outcomes at Bob and Eve are assumed to be
independent.
Therefore, the conditional secure success probability is determined as
\begin{equation}
\mathbb{P}\!\left(\mathcal{S}\mid \gamma_b,\gamma_e\right)
=
\big( 1 - \epsilon_b (\gamma_b) \big) \epsilon_e (\gamma_e).
\label{eq:secure_success_cond}
\end{equation}
Averaging over the SIRs $\gamma_e$ and $\gamma_b$, i.e., over the spatial randomness, fading, and random access activity, yields the
 secure success probability
\begin{equation}
P_{\mathrm{sec}}
\triangleq
\mathbb{P}(\mathcal{S})
=
\big( 1 - \epsilon_b \big) \epsilon_e,
\label{eq:Psec_exact}
\end{equation}
where $\epsilon_b$ and $\epsilon_e$ are previously derived in \eqref{eq:Peb_fbl} and \eqref{eq:Pee_fbl_closedform}, respectively.

This metric captures the joint impact of interference, network geometry, and FBL decoding on security performance.
Unlike the conventional secrecy outage probability, which is defined based on Shannon secrecy capacity in the infinite blocklength regime, $P_{\mathrm{sec}}$ captures decoding outcomes under FBL transmission, providing an operational measure of secrecy in short-packet communications.

\subsection{Finite Blocklength Secrecy Outage Probability}
\label{subsec:fbl_security_outage}

The FBL secrecy outage probability is defined as the probability that
secure success does not occur, i.e.,
\begin{equation}
P_{\mathrm{out}}
\triangleq
1 - P_{\mathrm{sec}}
=
1 - \big( 1 - \epsilon_b \big) \epsilon_e.
\label{eq:Pout_sec}
\end{equation}
This definition reflects an operational notion of secrecy that is well suited to
short packet transmission and contrasts with classical secrecy outage formulations based
solely on Shannon secrecy capacity \cite{Gopala2008,Zhou2011}.

\subsection{Secrecy Throughput}
\label{subsec:secrecy_throughput}

To jointly account for random access, FBL reliability, and secrecy, we
define the per-device secrecy throughput (bits/channel-use) as
\begin{equation}
T_{\mathrm{sec}}
\triangleq
p\,R\,P_{\mathrm{sec}}
=
p\,R\,\big( 1 - \epsilon_b \big) \epsilon_e.
\label{eq:Ts}
\end{equation}
This definition is consistent with throughput formulations in massive IoT networks, while
explicitly incorporating secrecy constraints \cite{tijana,Zhou2011}.

\section{Numerical Results}
\label{sec:results}

In this section, we present numerical results to evaluate the FBL secrecy performance of the considered uplink massive IoT network.  Unless otherwise stated, a baseline configuration assumes a BS with $G_b=15$ dBi
(i.e., $G_b=31.6$) and an IoT-class eavesdropper with $G_e=0$ dBi (i.e., $G_e=1$).

\begin{figure}[!b]
\centerline{\includegraphics[width=\linewidth,trim=95 268 100 280,clip]{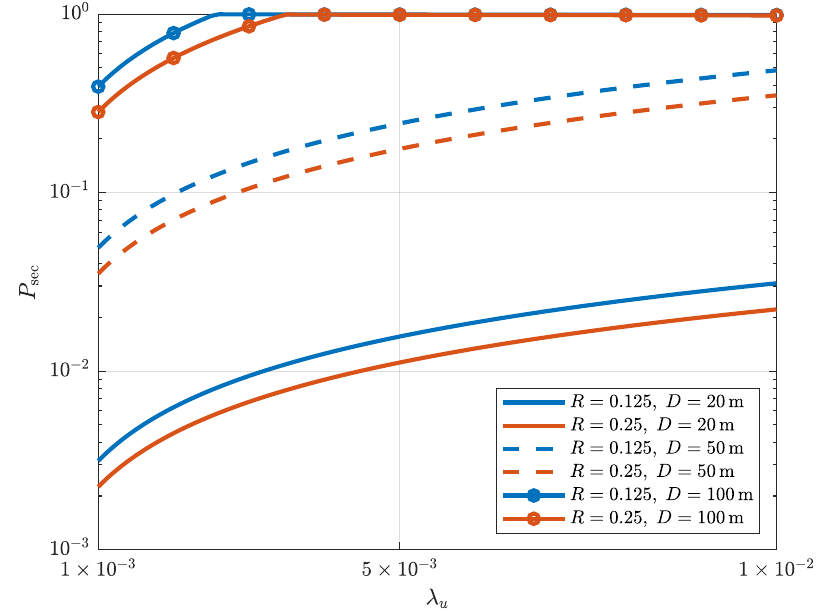}}
\caption{$P_{\mathrm{sec}}$ versus the IoT device density $\lambda_u$ for coding rates $R\in\{1/8,\,1/4\}$, distances $D\in\{20~\rm m,\,50~\rm m, 100~\rm m\}$, with a fixed blocklength $n=128$ and Nakagami parameter $m=1$. BS density is $\lambda_b=10^{-4}$, with activation probability $p        = 0.001$.}
\label{fig:Psec_lambda_u_D}
\end{figure}

Fig.~2 shows the secure success probability $P_{\mathrm{sec}}$
as a function of the IoT device density $\lambda_u$ for different coding rates
$R$ and eavesdropper radius $D$. 
As the IoT density increases, the secure success probability improves for all
considered parameter settings.
This behavior is due to the increased aggregate interference at the eavesdropper,
which degrades its decoding capability more severely than that of the legitimate
receiver.
Larger values of the eavesdropper radius $D$ further enhance secrecy performance,
since a wider region for the eavesdropper’s placement increases the likelihood that it is located farther from the IoT device, resulting in a lower received signal power at Eve.
For a fixed eavesdropper radius, lower coding rates yield higher values of $P_{\mathrm{sec}}$, reflecting the improved decoding reliability at the legitimate receiver under FBL transmission.
In particular, when $D=100\,\mathrm{m}$, the secure success probability approaches unity even at moderate IoT densities, indicating that the spatial separation of eavesdropper plays a dominant role in enhancing physical layer security in dense IoT networks, further highlighting that sufficiently large eavesdropper regions significantly mitigate interception capabilities.

Fig.~3 shows the secrecy outage probability $P_{\mathrm{out}}$ as a function of the BS antenna gain $G_b$ for different BS densities $\lambda_b$ and access probabilities $p$. It is observed that $P_{\mathrm{out}}$ decreases monotonically as $G_b$ increases for all considered network parameters. This behavior is due to the improved received signal strength at the BS with increasing antenna gain, which enhances the decoding reliability of the legitimate receiver. For a fixed $G_b$, a higher BS density results in a lower outage probability, as the effects of uplink interference are mitigated. Moreover, the curves corresponding to a smaller access probability  $p=10^{-2}$ exhibit lower outage levels compared to those with $p=5\times10^{-2}$, since reduced channel access alleviates aggregate interference. The performance gap among different values of $\lambda_b$ and $p$ becomes more pronounced at higher antenna gains, highlighting the dominant role of interference in determining secrecy outage performance.

Fig.~4 depicts the secure throughput $T_{\mathrm{sec}}$ derived in \eqref{eq:Ts} as a function of the access probability $p$ for different BS densities.  As $p$ increases, the secure throughput initially remains close to zero and then increases monotonically once reliable decoding at the BS becomes feasible. This threshold behavior is a consequence of FBL transmission, where the decoding error probability at the legitimate receiver remains close to one (e.g., $P_{\mathrm{sec}}$ and $T_{\mathrm{sec}}$ close to zero) until the effective SIR exceeds a minimum level.
For a fixed access probability,  higher BS density  improve the secure throughput by reducing the decoding error probability at the base station. 
These results highlight the importance of access control and network densification in maximizing secure throughput in massive IoT networks under FBL constraints.

\begin{figure}[!t]
\centering
\includegraphics[width=\linewidth,trim=95 268 100 280,clip]{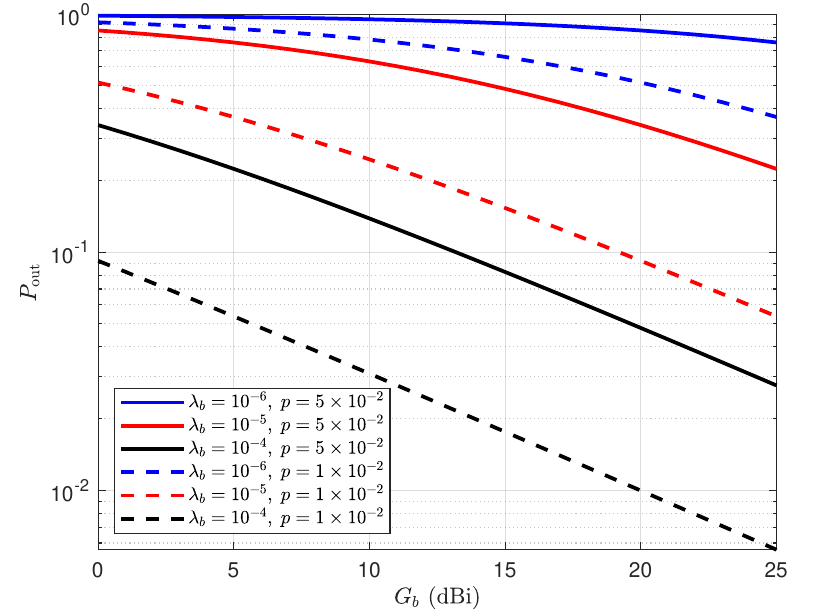}
\caption{$P_{\mathrm{out}}$ as a function of the BS antenna gain $G_b$ for three BS densities, $\lambda_b\in\{10^{-6},\,10^{-5},\,10^{-4}\}$, and two access probabilities, $p=5\times10^{-2}$ and $p=10^{-2}$. The coding rate is $R=1/4$, distance $D=100$ m, user density $\lambda_u=10^{-3}$, blocklength $n=128$, and Nakagami parameter $m=4$.}
\label{Fig_suc3}
\end{figure}

\begin{figure}[!t]
\centerline{\includegraphics[width=\linewidth,trim=95 268 100 280,clip]{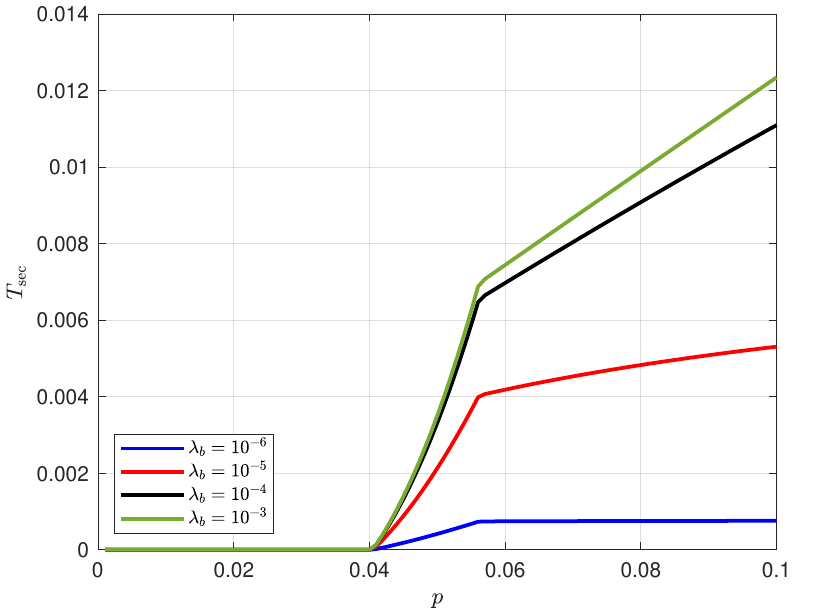}}
\caption{ $ T_{\mathrm{sec}}$ as a function of the access probability $p$, blocklengt $n=512$, and four BS densities, $\lambda_b\in\{10^{-6},10^{-5},10^{-4},10^{-3}\}$, rate $R=1/8$, distance $D=150$ m, user density $\lambda_u=10^{-3}$ and parameter $m=2$.}
\label{Fig_suc4}
\end{figure}


\section{Conclusion}
\label{sec:conclusion}
This paper presented a PLS analysis of massive uplink IoT networks operating under FBL constraints. By incorporating stochastic interference, Nakagami-\textit{m} fading and the spatial uncertainty of an eavesdropper, analytical expressions for key secrecy metrics, including the secure success probability, secrecy outage probability and secrecy throughput, have been derived. Numerical results illustrate a strong dependence of secure performance on  network density  and the relative proximity of the eavesdropper. These results confirm the validity of the analytical framework and provide useful insights for the design of secure short packet communication in dense IoT deployments. Future work may extend the framework to multi-eavesdropper scenarios or systems with adaptive transmission strategies.

\appendices
\section{}
\label{App1}
Based on linear approximation of the Q-function
 \cite[(13)]{tijana} and the parameters  $\mu$, $\theta$ and $a$ are  defined in Section III.B, the error probability of the eavesdropper can be approximated  as
\begin{equation} 
\begin{split}
 \epsilon_{e} &\approx \!\!\int_0^{D}\!\! \!\int_0^{\infty} \!\!\!\!\! \int_0^{\theta-a} \!\! \!\!\!\!f_{\Gamma_e}(\gamma_e|h_e,r_e\!)  f_{H_e\!} (h_e)   f_{R_e\!} (r_e) \mathrm{d}\gamma_e   \mathrm{d}h_e \mathrm{d}r_e\\ 
&\! + \!\!\int_0^{D} \!\!\int_0^{\infty} \!\!\int_{\theta-a}^{\theta+a} \!\! \bigg(\frac{1}{2} - \frac{\mu (\gamma_e - \theta)}{\sqrt{2\pi}}\! \bigg)  \\
& \times f_{\Gamma_e}(\gamma_e|h_e,r_e) f_{H_e} (h)    f_{R_e} (r_e)   \mathrm{d}\gamma _e \mathrm{d}h_e \mathrm{d}r_e \\
& = I_1 + I_2 - I_3 + I_4,
\end{split}
\label{eq31}
\end{equation}
where  the triple integrals in \eqref{eq31} are defined as follows
\begin{equation}
\begin{split}
\! \!  I_1 \! = \! \!  \int_0^{D}\!\! \!\int_0^{\infty} \!\!\!\!\! \int_0^{\theta-a} \!\! \!\!\!\!f_{\Gamma_e}(\gamma_e|h_e,r_e\!)  f_{H_e\!} (h_e)   f_{R_e\!} (r_e) \mathrm{d}\gamma_e   \mathrm{d}h_e \mathrm{d}r_e,
\end{split}
\label{I1}
\end{equation}
\begin{equation}
\begin{split} 
\!\!\!\!  I_2 \! = \! \! \frac{1}{2} \!\!\int_0^{D}\!\! \!\int_0^{\infty} \!\!\!\!\! \int_{\theta-a}^{\theta+a}  \!\! \!\!\!\!f_{\Gamma_e}(\gamma_e|h_e,r_e\!)  f_{H_e\!} (h_e)   f_{R_e\!} (r_e) \mathrm{d}\gamma_e   \mathrm{d}h_e \mathrm{d}r_e,
\end{split}
\label{I2}
\end{equation}
\begin{equation}
\begin{split}
\! \!  I_3 \! = \! \!  \frac{\mu}{\sqrt{2\pi}} \int_0^{D}\!\! \!\int_0^{\infty} \!\!\!\!\! \int_{\theta-a}^{\theta+a} \!\! \!\!\!\!\gamma_ef_{\Gamma_e}(\gamma_e|h_e,r_e\!)  f_{H_e\!} (h_e)   f_{R_e\!} (r_e) \mathrm{d}\gamma_e   \mathrm{d}h_e \mathrm{d}r_e,
\end{split}
\label{I3}
\end{equation}
\begin{equation}
\begin{split}
\! \!  I_4\! = \! \! \frac{\mu\theta}{\sqrt{2\pi}} \int_0^{D}\!\! \!\int_0^{\infty} \!\!\!\!\! \int_{\theta-a}^{\theta+a}  \!\! \!\!\!\!f_{\Gamma_e}(\gamma_e|h_e,r_e\!)  f_{H_e\!} (h_e)   f_{R_e\!} (r_e) \mathrm{d}\gamma_e   \mathrm{d}h_e \mathrm{d}r_e,
\end{split}
\label{I4}
\end{equation}
where $f_{R_e}(r_e)$,  $f_{H_e}(h_e)$ and $f_{\Gamma_e}(\gamma_e \!\mid \!h_e,r_e)$ are the  PDFs defined in \eqref{eq:fRe}, \eqref{eq:nakagami} and \eqref{pdfSINR}, respectively.

The integrals $I_1 $, $I_2 $ and $I_4$ have a similar form. After substituting \eqref{eq:fRe},  \eqref{eq:nakagami} and \eqref{pdfSINR} into \eqref{I1}, i.e., \eqref{I2} or \eqref{I4}, the following rules are applied: \cite[(8.251)]{grad}, 
\cite[(01.03.26.0004.01), (06.25.26.0006.01),  (07.34.21.0011.01),  (07.34.21.0011.01) and  (07.34.21.0084.01)]{wolfram}. The  integrals $I_1$, $I_2$ and $I_4$ are derived as 
\begin{equation}
\begin{split}
& \!\!\!\! I_1  \! = \!  \frac{ 2D }{\Gamma(m)\sqrt{\pi}\eta } \MeijerG*{2}{5}{5}{7}{\!-\frac{1}{4}, 0, \frac{1}{4},\frac{1}{2}, 1}{\frac{1}{2}, m, 0, -\frac{1}{2}, -\frac{1}{4}, 0, \frac{1}{4}\!\!}{ \frac{D^4t^2(\theta\!-\!a)m}{4G_e}},
\end{split}
\label{I1final}
\end{equation}

\begin{equation}
\begin{split}
\!\!\! I_{2} \!& =\!\frac{D}{\sqrt{\pi}\Gamma(m)\eta }\!\bigg(\!\!\MeijerG*{2}{5}{5}{7}{-\frac{1}{4}, 0, \frac{1}{4}\frac{1}{2},1}{\frac{1}{2}, m, 0, -\frac{1}{2}, -\frac{1}{4}, 0, \frac{1}{4}}{ \frac{D^4t^2(\theta\!+\!a)m}{4G_e}} \!\\
& - \MeijerG*{2}{5}{5}{7}{-\frac{1}{4}, 0, \frac{1}{4}\frac{1}{2},1}{\frac{1}{2}, m, 0, -\frac{1}{2}, -\frac{1}{4}, 0, \frac{1}{4}}{ \frac{D^4t^2(\theta\!-\!a)m}{4G_e}}\bigg),
\end{split}
\label{I2final}
\end{equation}

\begin{equation}
\begin{split}
\! I_{4} = & \frac{2 \mu \theta}{\Gamma(m)\pi 
 \sqrt{2}\eta} \bigg(\!\MeijerG*{2}{5}{5}{7}{\!-\frac{1}{4}, 0, \frac{1}{4}\frac{1}{2},1}{\frac{1}{2}, m, 0, -\frac{1}{2}, -\frac{1}{4}, 0, \frac{1}{4}\!}{ \frac{D^4t^2(\theta\!+\!a)m}{4G_e}\!} \\
& - \MeijerG*{2}{5}{5}{7}{-\frac{1}{4}, 0, \frac{1}{4}\frac{1}{2},1}{\frac{1}{2}, m, 0, -\frac{1}{2}, -\frac{1}{4}, 0, \frac{1}{4}}{ \frac{D^4t^2(\theta\!-\!a)m}{4G_e}}\!\bigg)\!.
\label{I4final}
\end{split}
\end{equation}

After substituting \eqref{eq:fRe},  \eqref{eq:nakagami} and \eqref{pdfSINR} into \eqref{I3}, and applying the rules \cite[(01.03.26.0004.01), (06.34.26.0005.01),  (07.34.16.0002.01)(07.34.21.0011.01),  (07.34.16.0002.01), and (07.34.21.0084.01)]{wolfram},
the integral $I_{3}$  is derived as 
\begin{equation}
\begin{split}
 I_{3} &=C_2\MeijerG*{2}{5}{5}{7}{-\frac{3}{4}, -\frac{1}{2}, -\frac{1}{4},0,-\frac{1}{2}}{-\frac{3}{2},0, m-\frac{1}{2}, -1, -\frac{3}{4}, -\frac{1}{2}, - \frac{1}{4}}{{y}_{2}}\\
 &-C_1\MeijerG*{2}{5}{5}{7}{-\frac{3}{4}, -\frac{1}{2}, -\frac{1}{4},0,-\frac{1}{2}}{-\frac{3}{2},0, m-\frac{1}{2}, -1, -\frac{3}{4}, -\frac{1}{2}, - \frac{1}{4}}{{y}_{1}},
\end{split}
\label{I3final}
\end{equation}
where $y_{1}$,
$y_{2}$, $C_{1}$ and $C_{2}$ are defined in Section III.B. After substituting  \eqref{I1final}, \eqref{I2final},  \eqref{I4final}  and 
\eqref{I3final} into \eqref{eq31}, final approximate FBL error probability is derived and given by {\eqref{eq:Pee_fbl_closedform}}.


\end{document}